\documentclass[a4paper,twoside]{article}
\newcommand{\affil}[1]{$^{\rm #1}$}
\usepackage[authoryear]{natbib}
\usepackage[usenames]{color}
\bibpunct{(}{)}{;}{a}{}{,}
\usepackage{graphicx}
\date{} 
\title{\large\bf\flushleft Reconstructing fossil sub-structures of the Galactic disk: clues from abundance patterns of old open clusters and moving groups}
\author{\parbox{\textwidth}{\flushleft
\vspace{-0.5cm}
{\it G.M. De Silva\affil{A}, K.C. Freeman\affil{B}, and J. Bland-Hawthorn\affil{C}}\\
\vspace{0.4cm}
{\small \affil{A}\,European Southern Observatory, Karl-Schwarzschild Str 2. D-85748 Garching, Germany.\\ Email:  gdesilva@eso.org}\\
{\small \affil{B}\,Research School of Astronomy and Astrophysics, Mount Stromlo Observatory, Australian National University. ACT 2611. Australia.}\\
{\small \affil{C}\,Institute of Astronomy, University of Sydney, NSW 2006, Australia. }}}
\begin{document}
\twocolumn[
\begin{changemargin}{.8cm}{.5cm}
\begin{minipage}{.9\textwidth}
\vspace{-1cm}
\maketitle
%
%
\small{\bf The long term goal of large-scale chemical tagging is to use stellar elemental abundances as a tracer of dispersed substructures of the Galactic disk. The identification of such lost stellar aggregates and the exploration of their chemical properties will be key in understanding the formation and evolution of the disk. Present day stellar structures such as open clusters and moving groups are the ideal testing grounds for the viability of chemical tagging, as they are believed to be the remnants of the original larger star-forming aggregates. Until recently, high accuracy elemental abundance studies of open clusters and moving groups having been lacking in the literature. In this paper we examine recent high resolution abundance studies of open clusters to explore the various abundance trends and reasses the prospects of large-scale chemical tagging.
}


\medskip
\medskip
\end{minipage}
\end{changemargin}
]
\small

\section{Introduction}
In order to follow the sequence of events involved in the formation of the Galactic disk, the critical components which need to be re-assembled are the ancient individual star-forming aggregates in the disk. But how do you reconstruct a star cluster which has dispersed a long time ago? Since the disk formed dissipatively and evolved dynamically, much of the dynamical information is lost. Any dynamical probing of the disk will only provide insights back to the epoch of last dynamical scattering. However, locked away within the stars, the chemical information may have survived the disk's dissipative history. Therefore the key to unraveling the epoch of dissipation is to investigate the chemistry of the disk stars and stellar structures.\\

The aim of chemical tagging \citep{fbh02} is to re-construct ancient star-forming aggregates, assuming such systems existed from an hierarchical aggregation formation scenario. Observationally the fact that stars are born in rich aggregates numbering hundreds to thousands of stars is supported by many studies from optical, infrared, millimeter and radio surveys \citep[eg.][]{carpenter,meyer,lada2}. The existence of stellar aggregates at earlier epochs are also observed in the form of old open clusters, stellar associations and moving groups. Further, theoretical hydrodynamical simulations indicate that star formation occurs in groups, where the original gas cloud undergoes fragmentation preventing contraction onto a single star \citep[eg.][]{jappsen,tilley,larson95}. Some clusters stay together for billions of years, whereas others become unbound shortly after the initial star-burst, depending on the star formation efficiency. If stars were not born in aggregates, it would be impossible to identify a given star's 
birth site. \\

While majority of stars that formed in clusters have dispersed into the field, the survival of old open clusters is interesting. It may be that these structures survived dispersion due
to a higher star formation rate and/or their Galactic orbits have avoided high
density regions, where interactions with giant molecular clouds are rare \citep{vandenbergh, finlay}. These old clusters are extremely important as they are the ideal probes for testing the historic conditions within a
cluster. The study of old open clusters will be the key to demonstrating the
existence of a unique chemical signature within star-forming aggregates. 

\section{ Open clusters}
Open clusters have historically been used for studying stellar evolution as all stars in
a given cluster are coeval. Their key attribute is that they provide a 
direct time line for investigating change. Young and old open clusters
are found in the disk, varying in age from several Myr to over
10 Gyr \citep[see compilation by][]{dias02}. The older ($>$ 1 Gyr) open clusters are less numerous than their younger counterparts and in general are more massive. These old open
clusters are excellent probes of early disk evolution.They constitute important fossils and are the likely left-overs of the early star-forming aggregates in the disk. Perhaps only the remnant cores are
left behind, while the outer stars have dispersed into the disk. \\

\subsection{Chemical homogeneity} \label{chemhom}
It is generally assumed that all stars in a cluster, born from the same parent proto-cluster cloud should contain the same abundance patterns. Theoretically, the high levels of supersonic turbulence linked to star formation in giant molecular clouds \citep{mckeetan02}, suggests that the interstellar medium of the gas clouds is well mixed and support the case for chemically homogenous clusters. Observational evidence show high levels of chemical homogeneity in open clusters. High accuracy differential abundance studies for large samples of Hyades open cluster F-K dwarfs, show little or no intrinsic abundance scatter for a range of elements \citep{paulson, desilva06}. Such internal homogeneity is also observed within old open clusters, e.g. Collinder 261 \citep{desilva07, carretta05}. Other old open cluster studies also support the case for internal abundance homogeneity \citep[e.g.][]{jacobson07, bragaglia08}, albeit for only a few stars per cluster. \\

The observations of homogeneity in old open clusters show that chemical information is preserved within the stars and effects of any external sources of pollution (e.g. from stellar winds or interactions with ISM) are negligible. Abundance differences may arise within cluster stars at various stellar evolutionary stages, e.g. due to mass transfer or internal mixing of elements during the dredge-up phases in giants. We do not expect that main sequence dwarf members of a cluster to show such effects. Further any internal mixing will only affect the lighter elements synthesized within the stars, while the heavier element abundances should remain at their natal levels. It is however interesting to note \citet{pasquini04}'s study on IC 4651 which show systematic abundance differences between the main sequence turn-off stars and the giants, although this maybe due to errors in deriving the stellar temperature scales. 

\subsection{Cluster sample}
Following the observational evidence discussed above (in section \ref{chemhom}), it is safe to assume internal abundance homogeneity holds for most open clusters in the disk. We now compare abundance patterns across different clusters. Mean cluster elemental abundances were taken from several high resolution abundance studies in the literature for clusters with ages greater than the Hyades. The list of clusters and their references are given in Table \ref{tab1}.  Figure \ref{iau_ocp1} plots the cluster mean abundances relative to Fe for the elements from Na to Zn. Figure \ref{iau_ocp2} plots the cluster mean abundances relative to Fe for heavier elements from Rb to Eu.

\begin{table}[h]
\begin{center}
\caption{Open cluster sample}\label{tab1}
{\scriptsize
\begin{tabular}{lcl}
\hline Cluster Name & [Fe/H] & Reference \\
\hline
NGC 6253& 0.46& \citet{carretta07} \\
NGC 6791 &0.47& \citet{carretta07} \\
NGC 7142& 0.08 & \citet{jacobson07} \\
NGC 6939 & 0.00 & \citet{jacobson07} \\
IC 4756      & -0.15 & \citet{jacobson07} \\
M 11 	  &0.10& \citet{gonzalez2000}  \\
NGC 2324  &-0.17& \citet{bragaglia08} \\
NGC 2477 &0.07&  \citet{bragaglia08}  \\
NGC 2660  &0.04&  \citet{bragaglia08} \\
NGC 3960  &0.02&  \citet{bragaglia08} \\
Be 32	    &-0.29&  \citet{bragaglia08} \\
NGC 6819 &0.09 & \citet{ngc6819} \\
NGC 7789 &-0.04&  \citet{tau05} \\
M 67		  &-0.03&  \citet{tau00} \\
NGC 2141 &-0.26& \citet{yong05}  \\
Be 31	 &-0.40& \citet{yong05}  \\
Be 29	 &-0.18& \citet{yong05}  \\
Be 20	 &-0.61& \citet{yong05}  \\
Tom 2	 &-0.45&  \citet{brown} \\
Mel 71 	 &-0.30&  \citet{brown} \\
NGC 2243 &-0.48&  \citet{gratton}  \\
Mel 66	 &-0.38&  \citet{gratton}  \\
Cr 261	 &-0.03& \citet{desilva07}  \\
Hyades	 &0.13& \citet{desilva06}  \\
\hline
\end{tabular}
}
\end{center}
\end{table}

\subsubsection{Errors and uncertainties}\label{errors}
Before the different clusters can be compared we stress that systematic differences are likely to exist due to differences in methodologies and scales. Since the studies are based on different clusters with no overlapping samples, such systematics are difficult to quantify. Where a study included an analysis of a reference star, such as the Sun, we can use the quoted differences as a guide to the expected systematic effects. In other studies, the reference solar abundance levels were simply adopted from past literature sources. This difference arising due to the solar reference value is $\approx$ 0.05 dex on average for all elements.\\

Systematic uncertainties differ per element based on how the individual elemental lines were analysed. Elements such as Na and Al are at risk of large and uncertain non-LTE effects, as well as modeling errors from unknown 3D hydrodynamical effects. We have adopted the published results based on standard LTE analysis to ensure a better comparison and any non-LTE analyses were not included. For the heavy neutron capture elements, errors due to hyperfine structures and isotope splittings can be significant. Even if these structures were taken into account in the original analyses, there may be differences in how these correction were applied between different studies. \\

The employed atomic line data varies between studies and gives rise to systematics. In differential analyses relative to the Sun or a reference star, the element line $gf$ values were recalculated. Other studies use laboratory measured $gf$ values from various literature sources. This may have large effects for certain elements, as noted by \citet{jacobson07}, whose Na abundances derived relative to Arcturus differs by as much as 0.2 dex from the other sources. However, they note that the Na abundances will be consistent with other studies if derived using laboratory based $gf$ values instead. \\

\begin{figure}[h]
\begin{center}
 \includegraphics[scale=0.38]{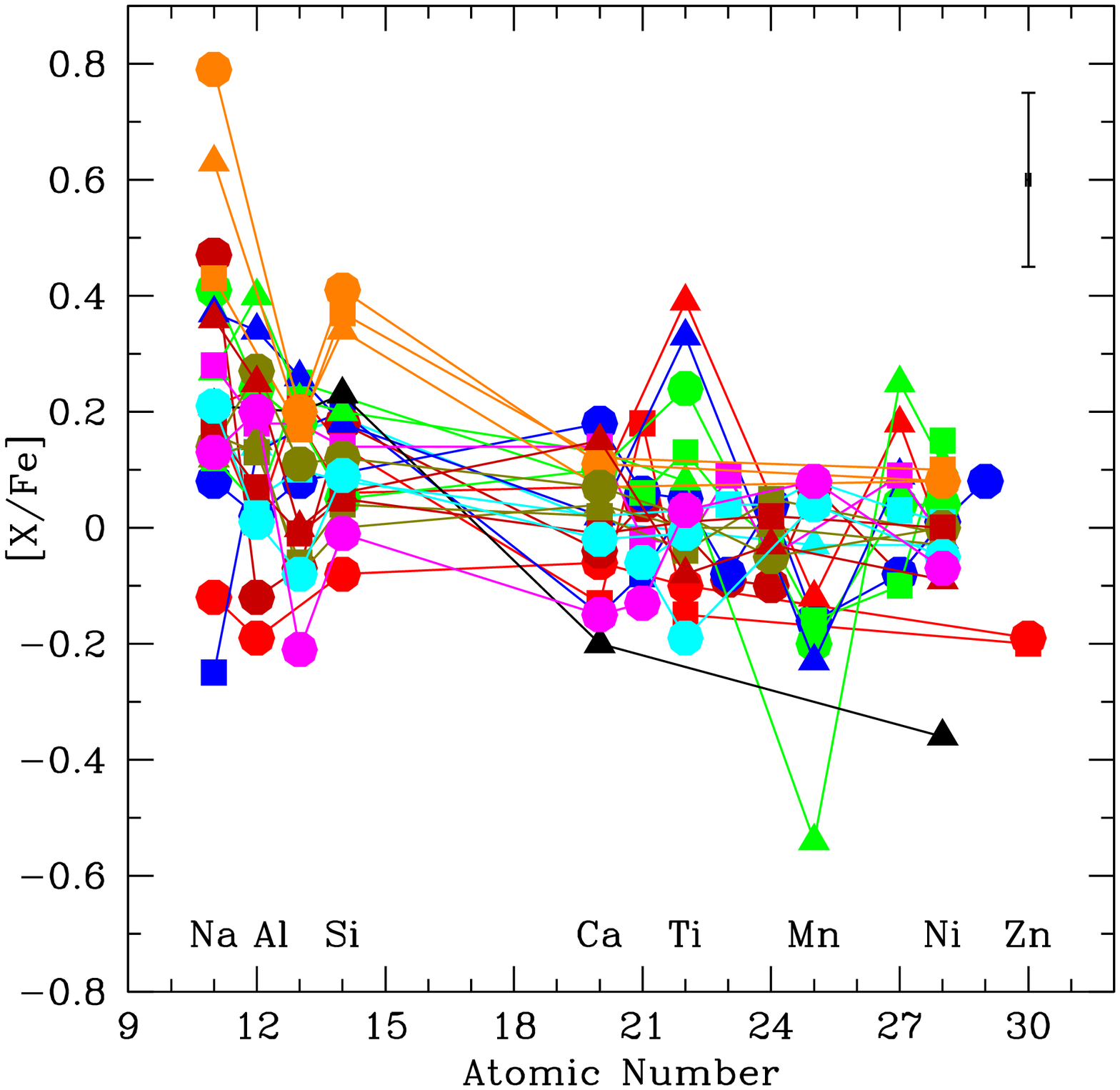}
  \includegraphics[scale=0.45]{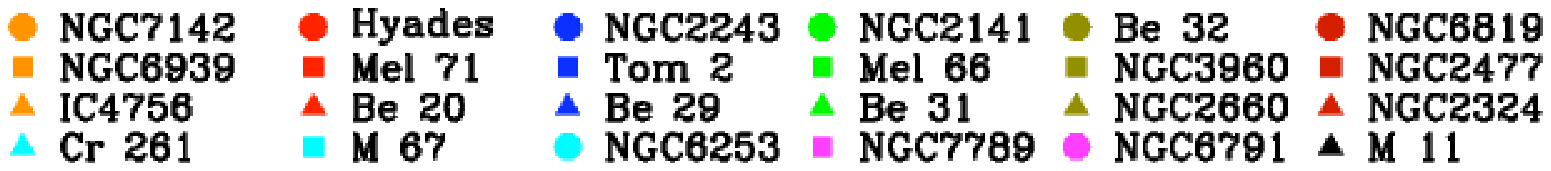}
  \caption{Abundances of old open clusters relative to Fe for elements from Na to Zn. Each symbol represents the mean abundance value for individual clusters. The error bars show the typical measurement error. Original references of the cluster data are given in Table \ref{tab1}. }
   \label{iau_ocp1}
\end{center}
\end{figure}

\begin{figure}[h]
\begin{center}
 \includegraphics[scale=0.38]{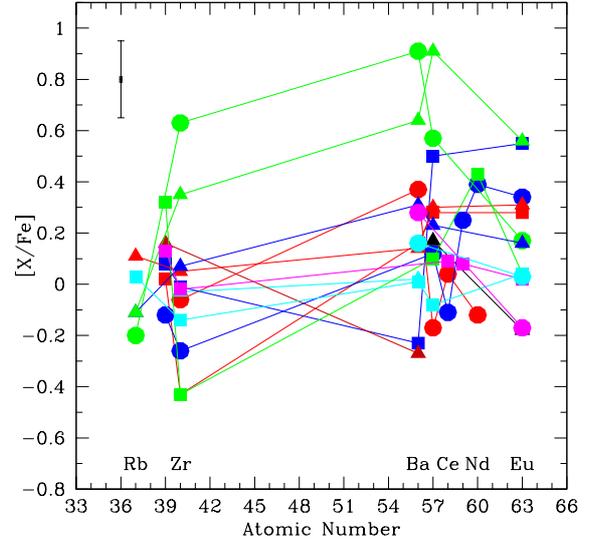} 
 \caption{Abundances of old open clusters relative to Fe for elements from Rb to Eu. The symbols are the same as in Figure \ref{iau_ocp1}.}
   \label{iau_ocp2}
\end{center}
\end{figure}

Other differences in analyses, such as the use of different model atmospheres and whether the abundance measurements are based on EWs or spectral synthesis also produces systematic  variations. Since much of these systematics cannot be accurately quantified for individual studies, we have not taken these into account, and plot the published mean abundance values in Figures \ref{iau_ocp1} \& \ref{iau_ocp2}. The error bars representing the typical measurement errors for the various elements in each study are shown. We refer the reader to the original studies for the individual measurement errors per element per cluster. Note the error bars do not account for other systematic uncertainties between different studies.\\

Further points to note include the number and type of stars in the different studies. The clusters studied by \citet{bragaglia08, carretta07} are based largely on 5-6 red clump stars per cluster, \citet{yong05} and \citet{jacobson07} studies are based on 2-5 red giants per cluster, while other studies are based on main sequence or turn-off dwarfs. As mentioned earlier in section \ref{chemhom}, we do not expect there to be any effects of stellar evolution for the heavier elements, while the abundances of lighter elements, such as Na, Al, Mg maybe affected in the giants by internal mixing. In this case they will no longer represent the clusters natal abundance levels, plus will be a source of scatter when comparing to dwarf members of other clusters. 

\subsection{Abundance signatures}

\begin{table}[h]
\begin{center}
\caption{The mean and standard deviation of cluster data for the individual elements.}\label{tab2}
\begin{tabular}{lcccc}
\\
\hline Element & $<$[X/Fe]$>$ & $\sigma$ \\
\hline
Na	&0.24 	&0.22 \\
Mg	&0.13	&0.15 \\
Al	&0.10	&0.13 \\
Si	&0.14	&0.12 \\
Ca	&0.02	&0.10 \\
Sc	&0.01	&0.10 \\
Ti	&0.04	&0.15 \\
V	&0.00	&0.08 \\
Cr	&-0.01	&0.05 \\
Mn	&-0.11	&0.19 \\
Co	&0.05	&0.12 \\
Ni	&-0.01	&0.10 \\
Cu	&0.08	&-\\
Zn	&-0.20	&0.01 \\
Rb	&-0.06	&0.12 \\
Y	&0.10	&0.15 \\
Zr	&-0.02	&0.30 \\
Ba	&0.21	&0.33 \\
La	&0.27	&0.31 \\
Ce	&0.01	&0.10 \\
Pr	&0.17	&0.12 \\
Nd	&0.23	&0.31 \\
Eu	&0.17	&0.24 \\
\hline
\end{tabular}
\end{center}
\end{table}

Figures \ref{iau_ocp1} \& \ref{iau_ocp2} show that different clusters have different elemental abundance patterns. There is also significant cluster to cluster scatter for many of the elements. We tabulate the mean abundance level and standard deviation per element in Table \ref{tab2}. As discussed above in section \ref{errors}, some of this scatter is likely to due systematic uncertainties, although there may be intrinsic variations, especially in elements showing excessive scatter, with $\sigma >$ 0.2 dex. We will now discuss the abundance variations in the open cluster sample in the context of exploring the various trends useful for chemical tagging. We will not discuss the individual clusters and their properties, as it is beyond our goals and can be found in the individual references in the literature.\\

Of the lighter elements, Na and Mg have the highest scatter, where most clusters are Na and Mg enhanced except for three clusters showing sub-solar levels. The other odd Z and alpha elements show cluster to cluster scatter within 0.15 dex with the average abundance being slightly super-solar. Enhanced alpha elements relative to Fe are indicative of a high rate of star formation where Type II SN dominate over the Type Ia SN. Therefore clusters showing such enhancement is likely to have undergone a phase of rapid star formation in comparison to those which show solar or sub-solar level alpha abundances.  Note for massive stars if internal mixing is effective, the Na and Mg could be produced internally and be more abundant than their natal levels. An anti-correlations of Na-O and Mg-Al would be a sign of such processes (as observed in globular clusters), but there is no evidence for such abundance patterns within open clusters.\\

As to why Na and Mg shows a larger scatter than other alpha elements remains a question; it is likely due to a different synthesis process that has not been fully understood. Note that only Si and Ca are considered to be the two pure alpha elements. Further of interest are the clusters that do not show equal alpha element abundance levels, e.g. with enhanced Si but deficient in Ca. This may represent some form of localized inhomogeneity unique to the time and site of the clusters' formation. Such abundance signatures will play a major role in large scale chemical tagging, when associating field stars to common origins.\\

The Fe-peak elements, thought to be produced via Type Ia SN, in general show the least scatter with the average abundance close to solar. This is expected given the abundances are plotted relative to Fe and we expect the Fe-peak elements to follow the Fe abundance. Nevertheless it is interesting to note that Ti has a larger scatter. Ti is considered the heaviest of the alpha elements, although by atomic number it falls into the Fe-peak group. It is not considered a pure alpha element either since Type Ia SN also contributes to its production in addition to the dominant Type II SN. Among the other Fe-peak elements Mn also shows a higher scatter, however this is dominated by a single cluster which is extremely Mn deficient. Note again that Mn could be synthesized in Type II SN as well as from Type Ia, where the yields are metallicity dependent \citep{mcwil03,shetrone}. From these examples it is clear that many of the elements behave differently within their groups. They do not necessarily vary in lock-steps and it is likely that various nucleosynthesis processes are at play.\\

The heavier s- and r- process elements show the largest scatter of all studied elements. Note the number of data points for these neutron capture elements are much less than for the lighter elements, a sign that they are difficult to measure. The mean abundance levels are super-solar for these elements. The exceptions are Zr with clusters showing both super and sub-solar Zr abundance levels, as well as Rb and Ce which has sub-solar abundances in the clusters although with only few data points. The s-process elements are synthesized under a low neutron flux environments, as in AGB stars and mixed into the ISM by stellar winds. The light s-process elements such as Zr seems to show a lower abundance compared to the heavier s-process elements such as Ba. Note however the opposite trend is observed for two clusters. Varying trends between Ba, La and Ce are also seen. The mostly r-process elements such as Nd and Eu, produced in high flux environments during Type II SN also show various trends among the open cluster abundances. Further the ratio of s- to r- process element abundance varies from cluster to cluster. We can expect this as both the s- and r-processes contribute at different levels to the production these neutron capture elements. Similar to the alpha and Fe-peak groups, this further demonstrates that the common group elements do not always vary in lock step. Their various abundance levels highlight the different conditions during the formation of the individual clusters. \\

The differences in the elemental abundance levels from cluster to cluster and the various element to element abundance trends seen among the clusters are important signs for large scale chemical tagging. It demonstrates that there is a sufficiently large abundance range to allocate $untagged$ stars into common formation sites. If the open cluster population showed to be chemically uniform, with little scatter and similar elemental abundance trends, then it will be difficult to chemically separate them. Therefore the present day cluster abundance variations is an important step forward for chemical tagging. In addition, the discussed decoupled nature of the elements further support large scale chemical tagging as it demonstrates a larger element parameter space in which to searching for chemical signatures. Higher accuracy, homogeneous abundance analyses will provide deeper views of the chemical substructure. This preliminary look however suggests that establishing cluster signatures with larger sets of elements is very much viable.

\section{Moving groups and stellar streams}
While the study of open clusters provide insight into the star-forming epochs of the Galactic disk, other structures exist in the disk that are of equal significance. These include moving groups, which are unbound groups of stars that share common motions around the Galaxy. These may also be relics of other processes, such as satellite accretion and mixing due to spiral arms, that have taken place in the
Galactic disk. \\

The concept of moving groups and superclusters was first advocated
by Olin Eggen in the 1960s.  Basically the stars form from a common
progenitor gas cloud. As the cluster orbits around the Galaxy, it
disperses into a tube-like structure around the Galaxy plane, and
after several galactic orbits, will dissolve into the Galaxy
background. The tube-like unbound groups of stars occupying extended
regions of the Galaxy were defined by  Eggen as superclusters. If
the Sun happens to be inside this tube structure,  the group members
will appear to be spread over the sky, but may be identified as a group
through their common space velocities. These group stars located within the solar neighborhood were defined as a moving group, and believed to be a subset of larger systems known as superclusters. \\

Besides dispersed open clusters, there are other manifestations of moving groups that have a dynamical orgin \citep{famaey}. Such dynamical streams are not thought to have originated from a dispersed open cluster, but are stars of different origins that have been swept up into a common orbit around the Galaxy by dynamical forces such as spiral density waves.  Many of these stellar streams have however not been subject to an abundance analysis, except for the Hercules stream \citep{fux, bensby07}. Despite sharing a common motion, the stars of the Hercules stream have different ages and chemistry. In fact the abundance trends of the member stars match that of the disk field stars.\\

\begin{figure}[h]
\begin{center}
 \includegraphics[scale=0.4]{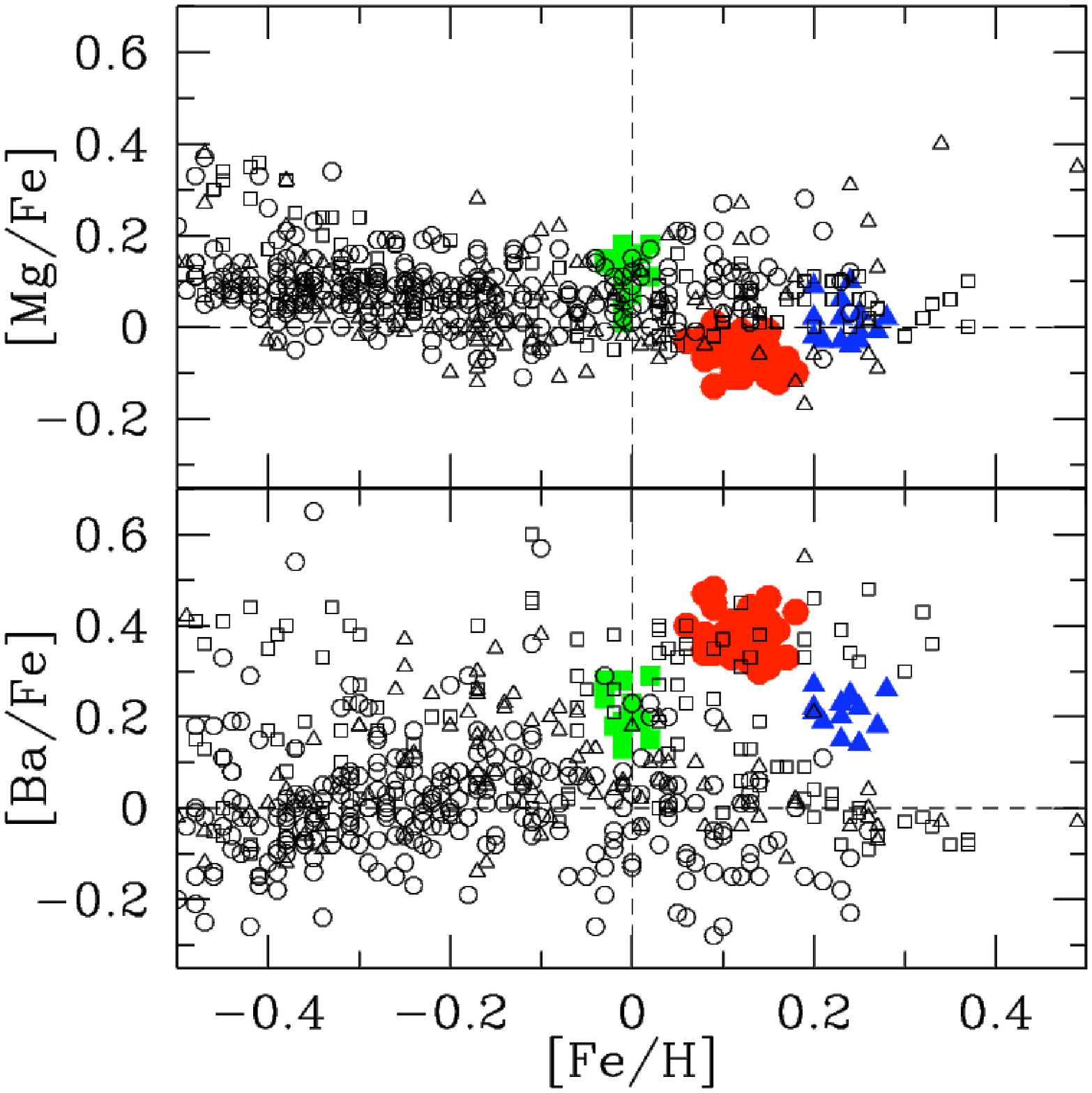} 
 \caption{Abundances of HR1614 moving group stars \citep[triangles]{hr1614} compared to the Hyades \citep[circles]{desilva06} and Collinder 261 \citep[squares]{desilva07} open clusters. The smaller open symbols represent background field stars \citep{reddy03, ap04, ed93}. The dotted lines mark the solar value.}
   \label{fig3}
\end{center}
\end{figure}

Conversely, an example of a true moving group is the HR1614 moving group \citep{e78a, e92, e98, fh}. The high levels of chemical homogeneity seen within the HR1614 moving group stars supports the case that it is a relic of an ancient star-forming event \citep{hr1614}.  Figure \ref{fig3} compares the abundances of the HR1614 moving group stars to that of the open clusters Hyades and Collinder 261. In contrast to the dynamically defined Hercules stream \citep[cf. Fig 3 in][]{bensby07}, it is clear that the HR1614 moving group in chemical abundance space represents a star cluster systems. Further, \citet{hr1614} show that chemical probing allows the differentiation between true members of the group and contaminating field
stars, which cannot be done with kinematics alone. The case of the nearby HR1614 moving group is surely not unique.  It is likely that there are other dispersed relic groups whose reality is not yet confirmed due to the lack of detailed chemical information. It is clear that kinematical
information alone cannot uncover the true story behind any dispersed stellar group in the Galactic disk. The presently demonstrated existence of a real relic moving group is a very important step and offers grand opportunities for chemical reconstruction of an bygone era.
\\

\section{Conclusion}
We have used high resolution elemental abundances of old open clusters from the literature to compare the cluster to cluster abundance trends for a large range of elements. We find that different clusters show different abundance levels for a given element, with some elements showing large scatter. Systematic uncertainties among the different studies is the source of much of the abundance scatter. However those elements showing a $\sigma > $ 0.2 dex is likely to be an indication of real cluster to cluster abundance variations. Further, various element to element abundance patterns were seen among the sample, highlighting the decoupled nature of the elements and the existence of chemical signatures unique to the clusters based on their time and site of formation. An $homogenous$ high resolution abundance study for a range of elements of the Galactic open cluster population \citep[e.g. the BOCCE project,][]{bocce} will be very insightful to further explore unique chemical signatures. \\

The internal homogeneity observed within open clusters, and the various abundance trends seen across the open cluster population, plus the fact that a dispersed moving group such as the HR1614 moving group is chemically identifiable, makes a solid step forward for the viability of large-scale chemical tagging. The next steps will be to chemically identify other dispersed stellar aggregates of the disk which may still retain some form of kinematical identity, e.g. the superclusters put forward by \citet{eggen78} which are though to be associated with open clusters and moving groups. The ultimate goal of large scale tagging is to reconstruct stellar aggregates that no longer retain any kinematical identify. High resolution multi-object spectrographs such as WFMOS on Gemini and HERMES on the AAT will be capable of obtaining the data needed for large scale chemical tagging of our Galactic disk.

\section*{Acknowledgments} 
G. De Silva would like to thank J. Sobeck for useful discussions regarding systematic uncertainties in the different abundance measurements. 



\end{document}